\documentclass[11pt,letterpaper]{article}
\usepackage{graphics,epsfig}
\usepackage[latin1]{inputenc}
\usepackage[T1]{fontenc}
\usepackage{amsmath}
\usepackage{amsfonts}
\usepackage{amssymb}
\usepackage{latexsym}
\usepackage[english]{babel}
\newcommand{\be}{\begin{equation}}
\newcommand{\ee}{\end{equation}}
\newcommand{\beq}{\begin{equation}}
\newcommand{\eeq}{\end{equation}}
\newcommand{\bea}{\begin{eqnarray}}
\newcommand{\eea}{\end{eqnarray}}
\begin{document}
\title{ Entanglement entropy for a Maxwell field: Numerical calculation on a two dimensional lattice}
\author{Horacio Casini\footnote{e-mail: casini@cab.cnea.gov.ar}, Marina Huerta\footnote{e-mail: marina.huerta@cab.cnea.gov.ar}\\{\sl Centro At\'omico Bariloche,
8400-S.C. de Bariloche, R\'{\i}o Negro, Argentina,} \\ {\sl Institute for Advanced Study, Princeton, NJ 08540, USA}}
\maketitle

\begin{abstract}
We study entanglement entropy (EE) for a Maxwell field in $(2+1)$ dimensions.  We do numerical calculations in two dimensional lattices. This gives a concrete example of the general results of our recent work \cite{gauge} on entropy for lattice gauge fields using an algebraic approach. To evaluate the entropies we extend the standard calculation methods for the entropy of Gaussian states in canonical commutation algebras to the more general case of algebras with center and arbitrary numerical commutators. We find that while the entropy depends on the details of the algebra choice, mutual information has a well defined continuum limit as predicted in \cite{gauge}. We study several universal terms for the entropy of the Maxwell field and compare with the case of a massless scalar field. We find some interesting new phenomena: An ``evanescent'' logarithmically divergent term in the entropy with topological coefficient which does not have any correspondence with ultraviolet entanglement in the universal quantities, and a non standard way in which strong subadditivity is realized. Based on the results of our calculations we propose a generalization of strong subadditivity for the entropy on some algebras that are not in tensor product. 
\end{abstract} 

\section{Introduction}
In a recent paper \cite{gauge} we have analyzed the problem of defining a local entropy for gauge fields. The inconveniences caused by the constraint equations of the physical degrees of freedom pointed to a natural setting within an algebraic approach for states and local algebras.  
 
The entropy on a region $V$ of the space is usually understood as the von Neumann entropy of the density matrix reduced to the degrees of freedom on that region. From the algebraic point of view, this is the entropy which results from the density matrix on a local algebra $\cal{A}_V$ associated to the region. In general, this algebra may have a center $\cal{Z}=\cal{A}_V \cap \cal{A}^{\prime}_V$, a set of operators that commutes both with the operators in the algebra and its commutant $\cal{A}^{\prime}_V$. Typically, the center is produced by the constraint equations.  Only the case with trivial center admits a bipartition of the Hilbert space as tensor product $\cal{H}_V \otimes\cal{H}_{\bar{V}}$ of subspaces of inner and outer degrees of freedom, and in this case the local entropy is an entanglement entropy for a global pure state. 

Of course, there is not a unique way to assign a local algebra to a region and different assignations give rise to ambiguities in the entropy. Even if these ambiguities are present in all theories, when the elementary excitations are not point like or more precisely, the operators are attached not to vertices but to links in the lattice, the standard prescription of identifying the region with the subset of operators attached to vertices within the region has to be revised. This is the case in lattice calculations for gauge fields, where the local gauge invariant algebra is generated by the electric link operators and Wilson loops along closed paths. The constraint equations give extra relations among the variables. A particular choice of local algebra with electric center has been discussed previously in the literature in a way unrelated to the algebraic formulation \cite{Polikarpov}.

In this paper we show how this general scheme applies in a specific example. We consider a Maxwell field theory in $2+1$ dimensions. In order to evaluate the entropies for general algebraic prescriptions for the local algebras we generalize the formulas for Gaussian states in canonical commutation algebras to the case where the the commutators are arbitrary matrices and the algebras have center. The techniques can be used for free (uncompactified) gauge fields (and more generaly massive or massless tensor fields) in any dimensions. We avoid using Wilson loop variables for this free model.   

We are able to show in detail the main prediction of \cite{gauge}: The large ambiguities in the entropy introduced by the uncertainties of the algebra choice and the universality of the mutual information in the continuum limit. This universality is a consequence of the fact that mutual information is ordered by inclusion of algebras. 
More technically, in the cases with center, we also show the classical Shannon term is not relevant for the continuum limit of the mutual information, and the calculation can be reduced to a unique arbitrary sector in the central decomposition.
 
Hence, the continuum limit eliminates the ambiguities in the relation between algebras and regions. In a certain sense, a geometric region has only meaning in terms of the content of the model once the continuum limit has been achieved. 

We also study some universal terms that can be obtained from the entropy itself, as the usual logarithmically divergent term due to the corners on the region boundary in $2+1$ dimensions. We find the logarithmic coefficient has also a curious additional contribution proportional to the number of connected components of the region. This should be regarded as related to the peculiarities of the gauge field in three dimensions. Surprisingly, this term is at the same time ultraviolet divergent and non local. However, we argue the ultraviolet nature of this term is not captured by any universal quantity in the model, i.e., the short  distance behavior of mutual information.  
 A related logarithmic term for the compactified Maxwell field in the limit of decompactification has been discussed in the literature in relation with the $F$-theorem \cite{headrick,kleba}.  

 This three dimensional model is dual to a ``truncated'' scalar field. The algebra generated by the electric and magnetic physical operators coincides with the one of time and space like derivatives of the scalar field, where the field operator  itself has been removed. We compare several universal terms of the gauge field (and truncated scalar) model with the model of a full scalar field. 

Even if with some specific choices of generating operators the local algebras have trivial center, the constraints reappear in other 
 interesting phenomena. For example, the usual geometric expression of strong subadditivity (SSA) property has to be reinterpreted in algebraic terms and generalized with respect to its usual form.  
 
We conclude with some discussion. In particular we revisit the issue of defining a topological entanglement entropy in a 
lattice using Levin and Wen \cite{wen} prescription in an algebraic way. This is connected with our discussion of strong subadditivity.

\section{Lattice Maxwell field in $2+1$ dimensions}
The physical operators $\bar{E}$ and $B$ of the $(2+1)$ dimensional Maxwell theory are written as
\bea
E_i&=&\partial_i A_0-\partial_0 A_i=F_{i0}\,,\\
B&=&\partial_1 A_2-\partial_2 A_1=\epsilon^{ij}\partial_i A_j=F_{12}\,,
\eea
in terms of the tensor field $
F_{\mu\nu}=\partial_{\mu}A_{\nu}-\partial_{\nu}A_{\mu}$. 
The canonical commutation relations (in Lorentz gauge)
\be
[A_{\mu}(x),\dot{A}_{\nu}(x^{\prime})]=ig_{\mu \nu}\delta^2(x-x^{\prime})\,,
\ee
give the gauge invariant commutation relations
\be
[B(x),E_j(x^{\prime})]=i\epsilon^{jk}\partial_k\delta^2(x-x^{\prime})\,,\hspace{1cm}[E_i(x),E_j(x^{\prime})]=[B(x),B(x^{\prime})]=0\,.
\ee
The Hamiltonian in terms of $E$ and $B$ is
\be
H=\frac{1}{2}\int dx_1 dx_2 (E^2+B^2)\,.
\ee

\begin{figure}[t]
\centering
\leavevmode
\epsfysize=4.3cm
\epsfbox{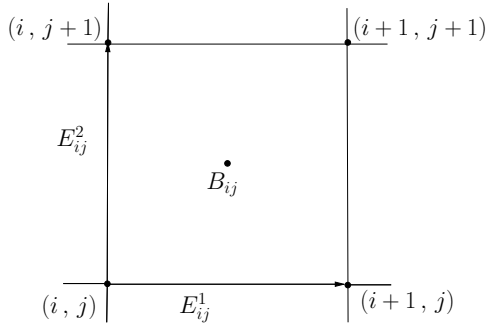}
\bigskip
\caption{The magnetic field is assigned to the center of the plaquette and the electric fields to the links.}
\label{figu1}
\end{figure}

Now, we discretize the model in a square lattice. The standard procedure for a gauge field is to assign the electric field variables to the links of the lattice and elementary Wilson loop operators to the plaquettes. For the non-compact Maxwell field we can consider directly the magnetic field operator (corresponding to the magnetic flux on the plaquette) and associate it to the dual lattice vertices in the middle of the plaquettes as shown in figure \ref{figu1}. Using directly the electric and magnetic variables allow us to profit from the Gaussianity of the model. 

More precisely, we define the electric operators $E^1,E^2$ associated to horizontal and vertical links respectively, as
$E^1_{(ij,i+1j)}$ and $E^2_{(ij,ij+1)}$, 
where $(ij,i^{\prime}j^{\prime})$ are the coordinates of the initial and final points of the link. This notation is useful but redundant since we can define the electric variables named by the initial vertex of the vector,
\bea
E^1_{ij}&\equiv& E^1_{(i j,i+1 j)}\,,\\
E^2_{ij}&\equiv& E^2_{(i j,i j+1)}\,.
\eea
The magnetic operator $B_{ij}$ is denoted by the left down corner 
$(i,j)$ of the plaquette (see figure(\ref{figu1})).

Hence, there are twice as many electric variables than magnetic variables. However, half the electric variables are redundant because of the constraint equations of electric flux (Gauss law) in two dimensions. This gives the relations 
\be
\sum_{b} E_{ab}=0\,,
\ee
where the sum is over all links $(ab)$ with common vertex $a$. In this equation, it is assumed that the electric field component is the corresponding one to the link direction and also that links have orientation which changes the sign of the electric field attached to it $E_{ab}=-E_{ba}$. 

In the lattice theory, the commutation relations become
\bea
\left[B_{ij},E^1_{(i^{\prime}j^{\prime},i^{\prime}+1j^{\prime})}\right]&=&-i \left(\delta_{i,i^{\prime}}\delta_{j,j^{\prime}}-\delta_{i,i^{\prime}}\delta_{j+1,j^{\prime}}\right)\,,\label{commem1}\\
\left[B_{ij},E^2_{(i^{\prime}j^{\prime},i^{\prime}j^{\prime}+1)}\right]&=&i \left(\delta_{i,i^{\prime}}\delta_{j,j^{\prime}}-\delta_{i+1,i^{\prime}}\delta_{j,j^{\prime}}\right)\,.
\label{commem}
\eea

Finally, the Hamiltonian writes
\be
H=\frac{1}{2}\left(\sum_v B_v^2+\sum_l E^2_l\right)\,,
\ee
where the sum is over the vertices for the magnetic variables and over the links for the electric ones. In contrast to the lattice Hamiltonian for a scalar field, this Hamiltonian is trivial as a bilinear form in the variables. All the dynamics is hidden in the constraint equations and non trivial commutation relations. 

\subsection{Maxwell-scalar field duality in the lattice}

In $(2+1)$ dimensions, the Maxwell theory is dual to the theory of the derivatives of a massless scalar $\phi$. The duality is written
\be
\partial_{\mu}\phi=\frac{1}{2}\epsilon_{\mu \nu \rho}F^{\nu \rho}\,,
\ee
giving the following identifications
\bea
\partial_0 \phi&=&B\,,\\
\partial_i\phi&=&-\epsilon_{ij}E_j\,.
\eea
This gives a complete one to one map between the theories, including the commutation relations and Hamiltonians. Note however, that the electromagnetic fields do not capture the full scalar theory but only the derivatives of the field. Hence the algebra of operators is strictly smaller than the one of the full scalar which includes $\phi$.

\begin{figure}[t]
\centering
\leavevmode
\epsfysize=6cm
\epsfbox{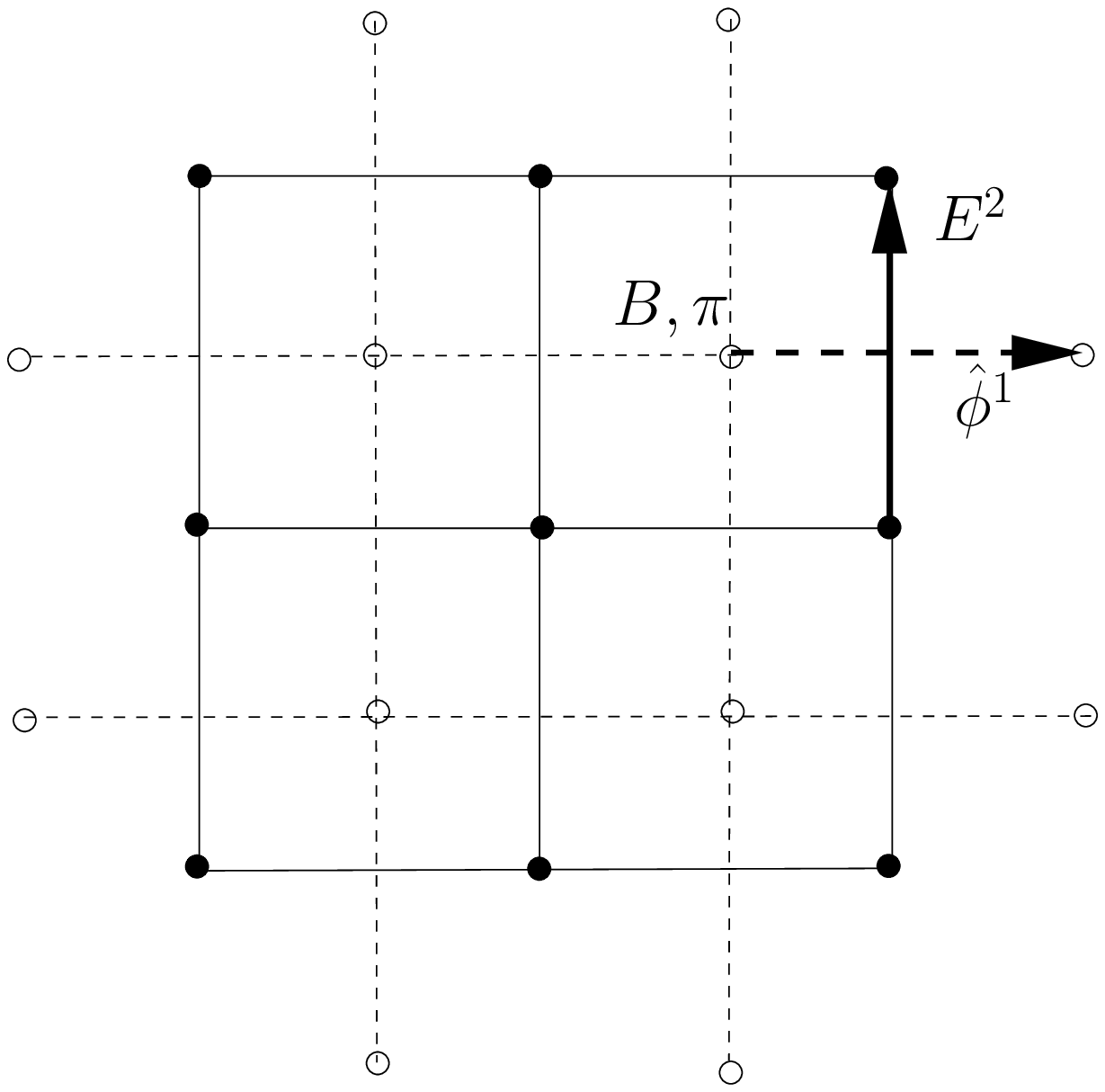}
\bigskip
\caption{Dual lattices: The magnetic field coincides with the momentum operator of the scalar field, and the electric field $E$ in some link is equal to a difference of scalar fields across the link in the dual lattice which is perpendicular to the one corresponding to $E$.}
\label{figu2}
\end{figure}

The discrete version of the above relations is expressed defining the scalar field variables on the sites of the dual lattice as shown in figure (\ref{figu2}).  The electric link operators are related to the differences of the scalar field operators in the orthogonal direction in the dual lattice
\bea
\phi_{\tilde{i}\tilde{j}}-\phi_{\tilde{i}\tilde{j}-1}&=&E^1_{(i j,i+1 j)}\,,\label{duality1}\\
\phi_{\tilde{i}-1\tilde{j}}-\phi_{\tilde{i}\tilde{j}}&=&E^2_{(ij,ij+1)}\,,
\label{duality}
\eea
and the magnetic operators are given by the corresponding momentum operators 
\be
B_{ij}=\pi_{ij}\,.
\ee 

\subsection{Truncated scalar theory}
\label{truncated}
The duality relations (\ref{duality1}) and (\ref{duality}) show it is equivalent to consider the gauge fields or the gradients of the scalar. We find convenient this later expression of the model for the entropy calculations.  Summarizing, this model consists of the subalgebra of the scalar field algebra (truncated scalar algebra) generated by 
\bea
\hat{\phi}^1_{i,j}&=&\phi_{i,j}-\phi_{i+1,j}\,,\\
\hat{\phi}^2_{i,j}&=&\phi_{i,j}-\phi_{i,j+1}\,,\\
\pi_{ij}&=&\dot{\phi_{ij}}\,.
\eea
The commutation relations 
\bea
\left[\hat{\phi}^1_{ij},\pi_{i^{\prime},j^{\prime}}\right]&=&i \left(\delta_{i,i^{\prime}}\delta_{j,j^{\prime}}-\delta_{i+1,i^{\prime}}\delta_{j,j^{\prime}}\right)\,,\\
\left[\hat{\phi}^2_{ij},\pi_{i^{\prime},j^{\prime}}\right]&=&i \left(\delta_{i,i^{\prime}}\delta_{j,j^{\prime}}-\delta_{i,i^{\prime}}\delta_{j+1,j^{\prime}}\right)\,,
\eea
are equivalent to (\ref{commem1}) and (\ref{commem}). The constraint equations for the electric field are mapped to the evident property
\be
\sum_{l\in p}\hat{\phi}_l=0\,,
\ee
where the sum is over the links $l$ on a plaquette $p$, with the same orientation along a curve encircling the plaquette.  

\begin{figure}[t]
\centering
\leavevmode
\epsfysize=6cm
\epsfbox{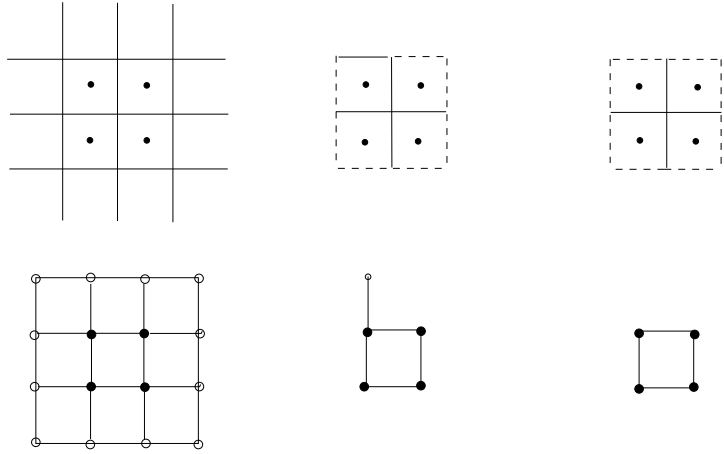}
\bigskip
\caption{ Some algebra choices for a square region. The upper three figures correspond to the gauge model and the ones at the bottom to the truncated scalar representation of the same algebras. Links with dashed lines mean the corresponding electric operator does not belong to the algebra. Marked dots correspond to magnetic operators in the algebra in the gauge model, and momentum operators in the scalar one. The left panel shows the electric center choice, where all electric and magnetic operators on the square belong to the algebra. Because of constraints the algebra also automatically contains the links coming out of the square, and there are more independent electric generators than magnetic ones. The central panel shows a trivial center choice, with balanced number of electric and magnetic degree of freedom. The panel on the right shows the magnetic center choice. Here, all electric operators on the boundary are missing and there is one more magnetic degree of freedom than the number of electric degrees of freedom.}
\label{figu3}
\end{figure}

The lattice Hamiltonian for the scalar field is
\be
H=\frac{1}{2}\left(\sum_{ij}\pi_{ij}+(\hat{\phi}^1_{ij})^2+(\hat{\phi}_{ij}^2)^2\right)\,.
\ee
From this Hamiltonian, the vacuum correlation functions for the field and momentum operators are found to be \cite{review}
\bea
f_{i,j}&=&\left<\phi_{00},\phi_{ij}\right>=\frac{1}{8\pi^2}\int_{-\pi}^{\pi}dx\int_{-\pi}^{\pi}dy\frac{\cos(ix)\cos(jy)}{\sqrt{2(1-\cos(x))+2(1-\cos(y))}}\,,\label{vacuumcor1}\\ p_{i,j}&=&\left<\pi_{00},\pi_{ij}\right>=\frac{1}{8\pi^2}\int_{-\pi}^{\pi}dx\int_{-\pi}^{\pi}dy\cos(ix)\cos(jy)}{\sqrt{2(1-\cos(x))+2(1-\cos(y))}\,.
\label{vacuumcor}
\eea
The correlators for the new variables $\hat{\phi}_{ij}$ and $\pi_{ij}$ can be easily written in terms of the ones of the $\phi$ variables, for example
\be
\left<\hat{\phi}^1_{00},\hat{\phi}^1_{ij}\right>=2 f_{i,j}-f_{i-1,j}-f_{i+1,j}\,.
\ee
These correlators are equivalent to correlators for lattice electric and magnetic fields and play an important role in the EE calculation we discuss later in Section 4.

\section{Local algebras and regions}
A ``region'' in the lattice has to be defined by the physical content of the model. That is, we must choose an algebra of local operators that defines it. In this sense, the assignation of algebras to regions is subject to ambiguities and several choices are possible. In the case the lattice operators are attached to vertices (such as a scalar field), a natural election seems to be to choose the local algebra associated to the region as the one generated by the operators attached to vertices within the region. Of course, this is just one possible choice. As we will discuss later, even in this case, we can think in different possibilities for the local algebra. 

In the case of gauge fields, we study three possible choices defined in \cite{gauge} which are shown in figure (\ref{figu3}): The algebras with electric, trivial, and magnetic center. Figure (\ref{figu3}) also shows the equivalent description of these algebras in terms of the dual truncated scalar algebra. 

In the electric center choice, we keep all the operators, inside and along the boundary. The constraints applied to the sites on the boundary show the electric fields normal to the boundary are automatically included in the algebra. As they commute with the rest of the operators on the algebra, they form the center. The electric center choice coincides with some constructions developed in the literature to define entropies in gauge theories \cite{Polikarpov}.

The trivial center case, consists in choosing all the electric and magnetic operators within the region and only one electric link operator along the boundary. This corresponds to remove the link operators along a boundary maximal tree \cite{gauge}. This election provides a good partition as tensor product between inside and outside degrees of freedom, giving place to a purely quantum entanglement entropy. However, this is not unique, we still have the possibility to vary the position of the electric field chosen at the boundary.

The magnetic center choice corresponds to the case where all the links along the boundary are removed. The center then, consists in the boundary Wilson loop, or equivalently the sum of the magnetic operators living in the interior plaquettes. 

For the truncated scalar algebra, we have analogue cases. The electric center, where the center is given by all the boundary links, the trivial center case, where the number of links and momentum operators are the same, and the magnetic case, with a one variable center given by the sum of all momentum operators on the region.

\section{Entropies of Gaussian states in terms of correlation functions}

In a general algebra, the center produces superselection sectors which cannot be changed by the local operators. The global state is then reduced into these sectors to give a block diagonal density matrix 
\be
\rho_{\cal A}=\left(
\begin{array}{ccc}
p_1 \,\rho_{{\cal A}_1} & & \\
 & \ddots& \\
 & & p_n \,\rho_{{\cal A}_n}
\end{array}
\right)\,.
\ee
 The entropy associated to the algebra ${\cal A}$ has a precise definition given by \cite{petz}
\be
S(V)=-\textrm{tr}(\rho_{{\cal A}}\log\rho_{{\cal A}})=H(\left\{p_k\right\})+S_Q({\cal A})\,,
\ee
where the first term corresponds to the classical Shannon entropy 
\begin{equation}
H(\left\{p_k\right\})=-\sum_k p_k \log(p_k)\,,
\ee
 of the probability distribution $\left\{p_k\right\}$ of the variables which simultaneously diagonalize the operators in the center. The second term $S_Q$ is an average of the corresponding purely quantum contributions 
 \be
 S_Q=\sum_k p_k \, S(\rho_{{A_k}})\,.
\ee
In the following, we are going to compute explicitly these entropies for the case of Gaussian states in algebras of coordinate and momentum operators. Hence, we consider cases with center containing operators with continuum spectrum (for example $q$) and the above formulas are generalized by converting the sum over discrete sectors $k$ into integrals.

\subsection{Algebra of canonical conjugated variables}
In order to calculate the EE for the models discussed, we need to generalize the method (see \cite{seealso,review}) for the case of Gaussian states and canonical conjugated variables $q_i,p_j$ with trivial center,
\be
\left[q_i,p_j\right]=i\delta_{ij}\,\,\,\,,\,\,\,\,\left[q_i,q_j\right]=\left[p_i,p_j\right]=0\,,
\ee
with $i,j\in V$. 
In this case, the entropy can be calculated in terms of the correlators 
\be
\left<q_i,q_j\right>=X^V_{ij}\,\,\,\,\,\,\textrm{and}\,\,\,\left<p_i,p_j\right>=P^V_{ij}\,,
\ee
as
\be
S_V=\textrm{tr} \left((\Theta+1/2)\log(\Theta+1/2)-(\Theta-1/2)\log(\Theta-1/2)\right)\,,
\label{realtimeapp}
\ee
where $\Theta=\sqrt{X^V.P^V}$, and $X^V$, $P^V$ are the correlators matrices (restricted to the algebra).

\subsubsection{Algebra of canonical conjugated variables with non trivial center}
\label{ntcenter}
The entropy of algebras with center formed by operators with continuous spectrum suffers from ambiguities due to the lack of a mechanism to fix the field normalization. However, mutual information between two algebras in tensor product (corresponding to two separated regions in a lattice model for example) \cite{gauge} is free from these ambiguities. Here, we deduce the general expressions for a set of coupled harmonic oscillators which we will use later in Sections 5 and 6 to calculate the mutual information between two sets for a scalar and a Maxwell field. 

Consider a set of harmonic oscillators with variables $q_i, p_i$, $i\in V=\{1,\dots,n\}$. We choose the algebra as the one generated by all the $q_i$ operators but only a subset of the momentum operators $p_i$ with $i\in B=\{k+1,...,n\}$. Hence this algebra has a center formed by the field $q_i$ with $i\in A=\{1,...,k\}$. 

We want to compute the entropy on this algebra for a state in $V$ that we assume is a Gaussian state. Then, it is convenient to write the density matrix in $V$ in a basis which simultaneously diagonalizes all elements in the center. In this case, we  choose the coordinate basis. We have for a Gaussian state
\be
\rho(q,q^\prime)=c\, e^{- \frac{1}{4}(q_i M_{ij}q_j+q_i^\prime M_{ij}q_j^\prime+2 q_i N_{ij}q_j^\prime)}\,,\label{made}
\ee  
where $c$ is a normalization constant, and due to hermiticity, $M$ and $N$ are real symmetric. The relation of these matrices with correlation functions on $V$ follows from
\be
\langle {\cal O}(q_i,p_j)\rangle =\int \prod_{i\in V} dq \,\, \left|{\cal O}(q_i,-i \partial_{q_j}) \rho(q,q^\prime)\right|_{q=q^\prime}\,.
\ee
We have
\bea
&&\langle q_i q_j\rangle\equiv X^{V}_{ij}= (M+N)^{-1}_{ij}\,,\\
&& \langle q_i p_j\rangle=\frac{i}{2}\delta_{ij}\,, \\
&&\langle p_i p_j\rangle\equiv P^{V}_{ij}=\frac{1}{4} (M-N)_{ij}\,.
\eea
All higher point functions are obtained by Wick's theorem for a Gaussian state. Inversely, we have
\bea
&&M=(2 X^V)^{-1}+2 P^V\,,\label{de}\\
&&N=(2 X^V)^{-1}-2 P^V\,.\label{ed}
\eea

Now, the probability density of a particular value $\tilde{q}^A$ for the variables $q^A$ on the center is again fixed by the correlation of the field in this region $A$,
\begin{equation}
{\cal P}(\tilde{q}^A)=\det((2\pi) X^A)^{-\frac{1}{2}} \,\,e^{- \frac{1}{2}\tilde{q}^A_i (X^A)^{-1}_{i j}\tilde{q}^A_j}\,.\label{ffi}
 \ee
The reduced density matrix in $B$ corresponding to this value of the variables on the center follows from (\ref{made}) by fixing these values for the fields $q, q^\prime\|_A=\tilde{q}^A$ on $A$, and a change on normalization, 
\be
\rho(q,q^\prime)=c^\prime\, e^{- \frac{1}{4}(q^B_i M^{BB}_{ij}q^B_j+2 \tilde{q}^A_i M^{AB}_{ij}q^B_j+q^{B \prime}_i M^{BB}_{ij}q^{B \prime}_j+2 \tilde{q}^{A}_i M^{AB}_{ij}q^{B \prime}_j+2 q_i^B N^{BB}_{ij}q_j^{B \prime}+2 \tilde{q}_i^A N^{AB}_{ij}q_j^{B \prime}+2  q_i^{B}N^{BA}_{ij}\tilde{q}_j^{A })}
\ee

Changing variables 
\bea
q^B&\rightarrow & q^B -(M^{BB}+N^{BB})^{-1} (M^{BA}+N^{BA}) \tilde{q}^{A}\,,\\
q^{B \prime}&\rightarrow &q^{B \prime} -(M^{BB}+N^{BB})^{-1} (M^{BA}+N^{BA}) \tilde{q}^{A}\,,
\eea
we get the density matrix 
\be
\rho(q,q^\prime)=c^{\prime\prime}\, e^{- \frac{1}{4}(q^B_i M^{BB}_{ij}q^B_j+q^{B \prime}_i M^{BB}_{ij}q^{B \prime}_j+2 q_i^B N^{BB}_{ij}q_j^{B \prime})}\,.\label{ggg}
\ee
Evidently, this change of variables does not change the entropy. Very conveniently, the density matrix (\ref{ggg})  is independent of the values of the field at the center. Hence, the average of the quantum entropy in $B$ over the values of the field $\tilde{q}^A$ on the center is trivial, and we get for the quantum part of the entropy
\bea
&&S_Q(V)=\textrm{tr} \left((\Theta+1/2)\log(\Theta+1/2)-(\Theta-1/2)\log(\Theta-1/2)\right)\,,\label{quant}\\
&& \Theta=\sqrt{\tilde{X}\tilde{P}}\\
&& \tilde{X}=(M^{BB}+N^{BB})^{-1}=(\left.X_V^{-1}\right|_B)^{-1} \,,\hspace{2cm}\tilde{P}=\frac{1}{4}(M^{BB}-N^{BB})=P_B\,.
\eea
The matrices $M^{BB}$ and $N^{BB}$ are just the restriction to $B$ of the matrices $M$ and $N$ which are in turn functions of the correlation functions in the region according to (\ref{de}) and (\ref{ed}).

The whole entropy $S(V)=S_Q(V)+H(A)$ contains also a classical Shannon term $H(A)$ due to the center probability distribution (\ref{ffi})
\be
H(A)=-\int (\Pi_{i\in A} dq_i) \, {\cal P}(\{q\}_A) \log({\cal P}(\{q\}_A))=  \frac{1}{2} \textrm{tr} \left(1+\log\left(2 \pi X^A\right)\right)\,.
\label{shannon}
\ee
This classical Shannon term can only have unambiguous meaning in relative entropy quantities, for example the relative entropy of two states or a mutual information between two regions for the same state. This is because the normalization of the fields in the center are not fixed by the commutation relations. For example, choosing the field $ q/\lambda $ instead of $q$ we get 
\be
\frac{1}{2} \textrm{tr} \left(1+\log\left(2 \pi X^A\right)-2\log(\lambda)\right)\,.\label{shannon1}
\ee 

The mutual information between two regions $V$ and $V^\prime$, with centers formed by the fields in $A\subseteq V$, $A^\prime\subseteq V^\prime$, is given by
\bea
I(V,V^\prime)&=&S(V)+S(V^\prime)-S(VV^\prime)=H(A,A^\prime)+I_Q(V,V^\prime)\nonumber \\
&=&\frac{1}{2}\textrm{tr} \left[\log(X^A)+\log(X^{A^\prime})-\log(X^{AA^\prime}))\right]+S_Q(V)+S_Q(V^\prime)-S_Q(VV^\prime)\,.
\label{mutual}
\eea 
This gives the desired expression for the mutual information of two algebras with center purely in terms of the correlation function matrices. 

\subsection{Generalization for arbitrary commutators and constraints}
 In the case the variables satisfy canonical commutation relations and the algebra has a trivial center, the entanglement entropy associated to a region $V$, can be calculated in terms of the correlators restricted to $V$ according to eq. (\ref{realtimeapp}). In the case the algebra has a non trivial center the entropy is given by the sum of the quantum (\ref{quant}) and classical (\ref{shannon}) parts. In this section, we show how these formulas can be extended to the case of  conjugated variables having general numeric commutators. This is the case of the physical variables $E,B$ in Maxwell theory, or the variable $\hat{\phi}$, $\pi$ in the truncated scalar model.
 
Consider an operator algebra with non canonical commutation relations
 \begin{equation}
\left[q_i,p_j\right]=iC_{ij}\,,
\label{corr}
\end{equation}
and correlators
\begin{eqnarray}
\left<p_i,p_j\right>&=&P_{i,j}\\
\left<q_i,q_j\right>&=&X_{i,j}\\
\left<q_i,p_j\right>&=&\frac{i}{2} C_{ij} \label{cor}\,.
\end{eqnarray}
Suppose we are interested in a subalgebra without center $\{q_i,p_i\}$ with $i\in V$. We can define new canonical variables $\hat{q}_i$, $p_j$, using 
\begin{equation}
\hat {q}_i=({C^V})^{-1}_{ik} q_k\,.\label{transf}
\end{equation}
The correlation functions restricted to the region will be
\begin{eqnarray}
\left<\hat{q}_i,\hat{q}_j\right>&\equiv&(C^V)^{-1} X^V\left((C^V)^{-1}\right)^T\,,\\
\left<\hat{q}_i,p_j\right>&=&\frac{i}{2}\left.\delta_{ij}\right|_V\,,\\
\left<p_i,p_j\right>&\equiv& P^V\,.
\label{redcorr}
\end{eqnarray}
The entropy is then calculated in terms of $\Theta=\sqrt{\left<\hat{q},\hat{q}\right>.\left<p,p\right>}$ as (\ref{realtimeapp}).

\begin{figure}[t]
\centering
\leavevmode
\epsfysize=5cm
\epsfbox{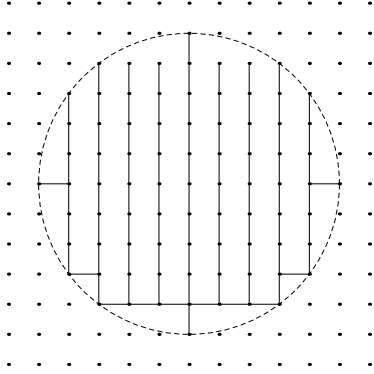}
\bigskip
\caption{A circle on a square lattice. A maximal tree of links inside the region gives all linearly independent link variables in the truncated scalar model. We keep only variables in an arbitrary maximal internal tree to make actual computations. Analogously, in the gauge model we have to keep only the electric fields that are orthogonal to this tree in the dual lattice.}
\label{arbol}
\end{figure}

It is important to notice that the entropy is a function of the algebra and the global state. Hence, if instead of the variables $q_i$, $p_j$, we take arbitrary linear combinations of these, we end up with the same entropy, as long as we consistently change the correlation matrices and commutators. The same can be said for the case where there are constraints. For example, we can have more variables $q$ than $p$ because some combinations of the $q$ variables are zero, as happens for the electromagnetic field and the truncated scalar.\footnote{Redundant variables appear also if we choose to express the entropies in terms of correlation functions in spacetime (as opposite to space) as was recently proposed \cite{sorkin}. In this case, the equations of motion play the role of the constraints on the operators of the algebra.}  We can eliminate the redundant variables in many different ways and keep an equal number of coordinate and momentum variables, and then compute the entropy as described above. The entropy is invariant under these ``gauge fixings''. See figure \ref{arbol}.  

\subsubsection{The case with center}
The general case in gauge theories involves algebras with center and non trivial commutators. The previous discussion about the entropy can be generalized to this case.
Consider the algebra generated by $q_i, p_j$, with $i\in V=\{1,\dots,n\}$ and $j\in B=\{k+1,...,n\}$ with $B\subset V$. We assume $[q_i,p_j]=0$ for $i\in A=\{1,...,k\}$, $j\in B$, in such a way that $q_i$, $i\in A$ span the center of the algebra. Using again the transformation (\ref{transf}) we arrive at the case studied in  (\ref{ntcenter}) and we find the quantum contribution to the entropy is given by
\bea
&&S_Q(V)=\textrm{tr} \left((\Theta+1/2)\log(\Theta+1/2)-(\Theta-1/2)\log(\Theta-1/2)\right)\,,\\
&& \Theta=\sqrt{\tilde{X}\tilde{P}}\\
&& \tilde{X}=(\left.X_V^{-1}\right|_B)^{-1}= (C_{VB}^TX_V^{-1}C_{VB})^{-1}\,,\hspace{2cm}\tilde{P}=P_B\,.
\eea
Here $C_{VB}$ is the commutation matrix  (\ref{corr}) between $q_i$ with $i \in V$ and $p_j$ with $j \in B$.
The classical contribution has the same form as before
\be
H(A)=\frac{1}{2} \textrm{tr} \left(1+\log\left(2 \pi X_A\right)\right)\,.
\label{class}
\ee
The case for a center formed by $p_i$ with $i\in A=\{1,...,k\}$ is analyzed in the same way, interchanging $P\leftrightarrow X$.

\subsection{Correlators for the vacuum state}
We are interested in vacuum entropies. Here, we show how to compute the correlators for simple quadratic Hamiltonians relevant for the Maxwell field. 

The vacuum correlators for Gaussian states can be directly calculated from the kernel of the quadratic Hamiltonian \cite{review} for the free scalar field. It is easy to show that this result can be generalized for the case of variables with non canonical commutation relations.
Consider a theory with Hamiltonian
\begin{equation}
H(q,p)=\frac{1}{2}\left( \sum_i {p_i} ^2+\sum_{i,j} q_i M_{ij} q_j\right)\,,
\label{hamiltonian}
\end{equation}
for the canonical conjugated variables $\hat{q}$, p, with 
\begin{equation}
\hat {q}_i=({C})^{-1}_{ik} q_k\,,
\end{equation}
and $C$ defined in (\ref{corr}). Changing variables, the Hamiltonian takes the form
\begin{equation}
H(\hat{q},\pi)=\frac{1}{2}\left( \sum_i {p_i}^2+\sum_{i,j} \hat{q_i} \hat{M}_{ij} \hat{q}_j\right)\,,
\end{equation}
where
\begin{equation}
\hat{M}=C^T.M.C\,.
\end{equation}

The two point correlation functions for the fundamental state are given in terms of $C$ and $M$ as \cite{review},
\begin{eqnarray}
\left<p_i,p_j\right>&=&\frac{1}{2}\left(\hat{M}^{1/2}\right)_{ij}\,,\\
\left<\hat{q}_i,\hat{q}_j\right>&=&\frac{1}{2}\left(\hat{M}^{-1/2}\right)_{ij}\,,\\
\left<\hat{q}_i,p_j\right>&=&\frac{i}{2}\delta_{ij}\,.
\end{eqnarray}
This gives for the original variables
\begin{eqnarray}
\left<p_i,p_j\right>&=&\frac{1}{2}\left(\left(C^T.M.C\right)^{1/2}\right)_{ij}\,,\\
\left<q_i,q_j\right>&=&\frac{1}{2}\left(C.\left(C^T.M.C\right)^{-1/2}.C^T\right)_{ij}\,,\\
\left<q_i,p_j\right>&=&\frac{i}{2}C_{ij}\,.
\end{eqnarray}
In the case $M$ is the identity matrix, which is relevant for the Maxwell field, we have
\begin{eqnarray}
\left<p_i,p_j\right>&=&\frac{1}{2}\left(\left(C^T.C\right)^{1/2}\right)_{ij}\,,\label{dfg1}\\
\left<q_i,q_j\right>&=&\frac{1}{2}\left(C.\left(C^T.C\right)^{-1/2}.C^T\right)_{ij}=\frac{1}{2}\left(C.C^T\right)^{1/2}_{ij}\,,\label{dfg2}\\
\left<q_i,p_j\right>&=&\frac{i}{2}C_{ij}\,.
\label{correlators}
\end{eqnarray}

In section (\ref{truncated}) we showed how correlation functions for the Maxwell field in $2+1$ dimensions are obtained from scalar correlation functions. Of course, this coincides with the above formulas when applied directly to the gauge field. These formulas can be used to obtain the lattice correlators of electric and magnetic fields in other dimensions. Notice that 
formulas (\ref{dfg1}) and (\ref{dfg2}) for the correlators do not have singularities for non invertible correlator matrix.
 This means we can use them for the gauge fields directly without necessity of solving for the constraints. Indeed we have $M=1$ for the Maxwell Hamiltonian expressed in the variables $E$ and $B$ where the constraints have not been used. Then, the commutator matrix is in general rectangular, but this does not affect the validity of (\ref{dfg1}) and (\ref{dfg2}).

\section{Some examples with a massless scalar field}

\begin{figure}[t]
\centering
\leavevmode
\epsfysize=6.5cm
\epsfbox{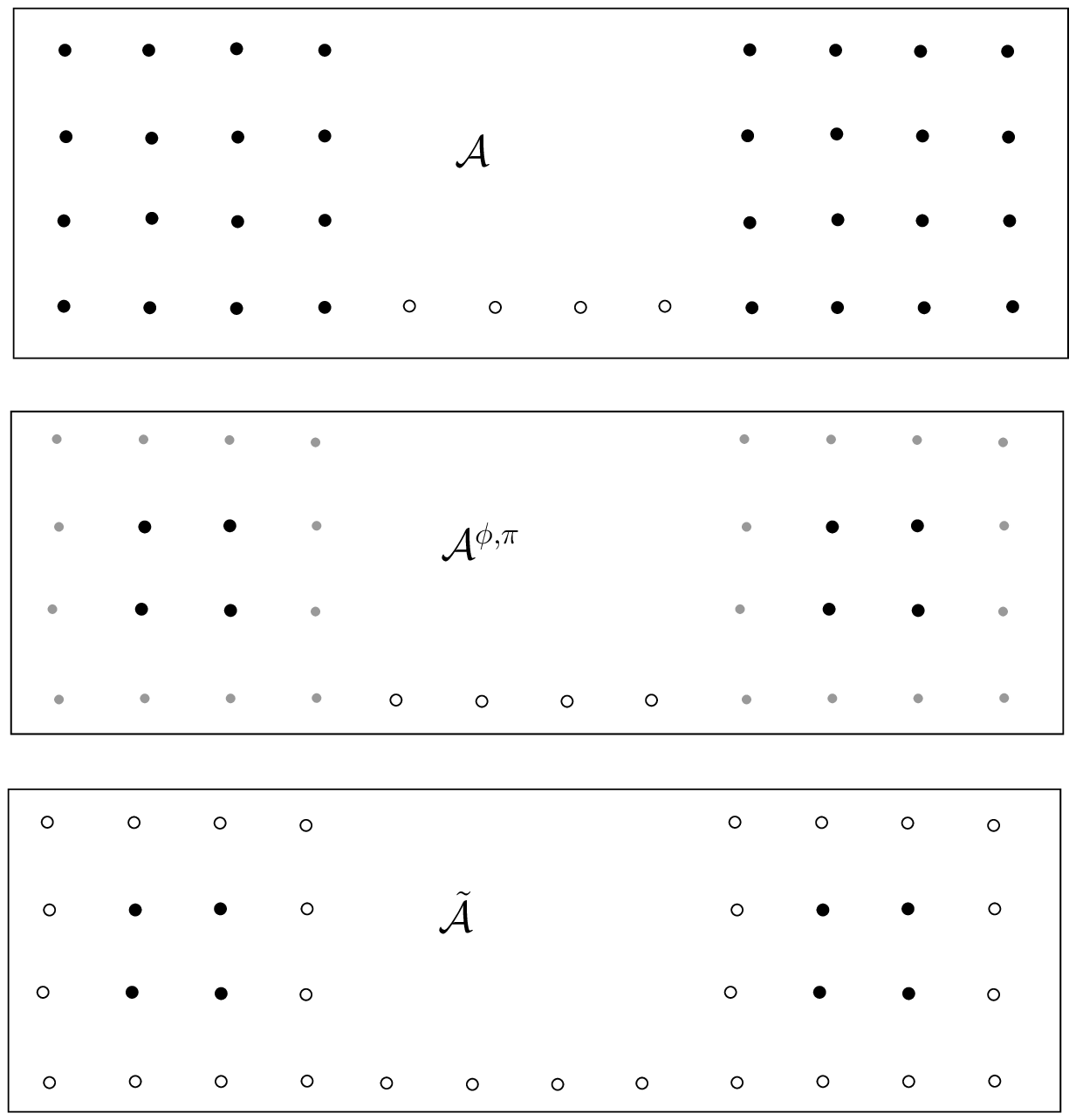}
\bigskip
\caption{We compute the mutual information for two squares with different algebra choice. Upper panel: Trivial center, where operators $\phi_{i}$ and $\pi_{i}$ are attached to each vertex (black dots). Empty dots are shown with the purpose to describe the position of the squares in the lattice and no operator are attached to them. Middle panel: Non trivial center. Operators $\phi_{i}$ and $\pi_{i}$ are attached to black vertices. At gray vertices the corresponding $\pi$ operators are removed in ${\cal A}_\phi$ and the $\phi$ operators are removed in ${\cal A}_\pi$. The center is generated by operators remaining at gray vertices. Lower panel: The algebra $\tilde{{\cal A}}$ is the full algebra of the central square of size $n-2$ points.}
\label{figu5}
\end{figure}

Before considering the gauge field and the truncated scalar, we exemplify the methods discussed above with a massless scalar fiels and several different algebra choices.

We consider a simple case of two square regions $V$ and $W$ of size $n$ lattice points, separated by the same number $n$ of lattice sites, and compute the mutual information for four different algebras for each square as shown in figure (\ref{figu5}): 

\noindent a) The  full algebra ${\cal A}$ of the squares of size $n$ with trivial center, already studied in \cite{ang}. 

\noindent b) The algebra ${\cal A}^\phi$ which results by removing all the $\pi_i$ operators from the boundary with a center formed by the remaining $\phi_i$ along the boundary. 

\noindent c) The opposite case where the removed fields are the $\phi_i$ and we have an algebra ${\cal A}^\pi$ with momentum center. 

\noindent d) The algebra $\tilde{{\cal A}}$ resulting from the elimination of all operators of the boundary (that is, we consider squares of side $n-2$ in the center of the original squares). 

We have for these algebras 
\be
{\cal A}\supset{\cal A}^{\phi} \,,\,{\cal A} ^{\pi}\supset \tilde{\cal A}\,.
\ee
The mutual information is monotonously increasing with the algebra. Hence, we expect to have
\be
I_{{\cal A}}(V,W)\ge I_{{\cal A} ^{\phi}}(V,W),\,\,\,I_{{\cal A} ^{\pi}}(V,W)\ge I_{\tilde{\cal A}}(V,W)\,,\label{obey}
\ee
where $V$ and $W$ are the two squares.

In figure (\ref{figu6}) we show the numerical calculation of the mutual information between two squares of the same size $n$ and separated by a distance $n$, being $n$ the number of vertices, for the different algebra choices. Here, we use the correlators (\ref{vacuumcor1}), (\ref{vacuumcor}) and formula (\ref{mutual}) to calculate the mutual information. 
The figure shows the ordering relations (\ref{obey}) are obeyed. 
We expect a convergence of the mutual information for large $n$ to the continuum limit. In fact, the limit values $I_0$ of the mutual information obtained by a fit of the form $I(V,W)=I_0+I_1 n^{-1}+I_2 n^{-2}+I_3 n^{-3}$ are $0.03299$, $0.03302$, $0.03304$ and $0.03308$ for the algebras ${\cal A}$, ${\cal A}^{\phi}$, ${\cal A}^{\pi}$ and $\tilde{\cal A}$, respectively, showing a remarkably fast convergence to a common constant value already for sets of size $n=35$. This mutual information is a very small number, approximately $1/20$ bit for infinitely many degrees of freedom in the continuum limit. This reflects the locality of the theory. 

Figure (\ref{figu6}) illustrates our general argument \cite{gauge} on why mutual information must have a unique continuum limit disregarding the details of the algebra choice for the region. This is because some prescription for algebra choice (in this case ${\cal A}^{\phi}$ and ${\cal A}^{\pi}$) can be bounded above and below by another prescription with slightly different size for the regions (in this case ${\cal A}$ and $\tilde{\cal A}$). As the continuum limit is reached, necessarily, all these prescriptions lead to the same values.

\begin{figure}[t]
\centering
\leavevmode
\epsfysize=5cm
\epsfbox{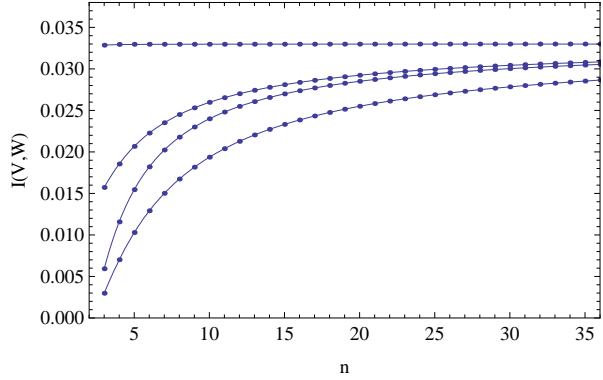}
\bigskip
\caption{Mutual information between two square regions of size $n$ and separation distance $n$ for four different algebra choices, from top to bottom ${\cal A}$, ${\cal A}^{\phi}$, ${\cal A}^{\pi}$ and $\tilde{\cal A}$. 
The curves corresponding to algebras with center ${\cal A}^{\phi}$ and ${\cal A}^{\pi}$ are in between the ones for the full algebras ${\cal A}$ and $\tilde{\cal A}$. Fitting the curves with an expansion in inverse powers on $n$ all four curves lead to the same $n\rightarrow \infty$ limit $I(V,W)\simeq 0.0330$.}
\label{figu6}
\end{figure}

On the other hand, regarding the entropy, different choices of local algebra result in dramatic changes. In $(2+1)$ dimensions, we expect the entropy for massless theories and polygonal sets to have the following form as a function of the overall size $n$,
\be
S=c_1 n +c_{\log}\log n-c_0+{\cal O}(n^{-1})\,.
\ee
 The coefficient $c_{\log}$ is universal and comes from a sum over the corners $v$ of the region \cite{ang,ang1},  and for each corner it depends on the vertex angle $\theta_v$,
 \be
 c_{\log}=-\sum_v s(\theta_v)\,.
 \ee
 For a square region and a massless scalar field, we have

\noindent a) Trivial center - full algebra
	\bea
 c_{\log}&=&-0.0472\,\Rightarrow\,s(\pi/2)=-\frac{c_{\log}}{4}=0.0118\,,\label{yapeyu}\\ 
	c_1&=&0.309,\hspace{1cm}  c_0=-0.0881 \nonumber\,,
	\eea
where we have allowed for an additional $n^{-1}$ term in the fit and taken squares of size up to $n=35$.
The presence of a center and in consequence, a classical contribution to the total entropy, results in relevant changes on the constant and area terms while the only preserved term is the logarithmic one. This can be seen in the coefficients we find for two different center choices,

\noindent b) Center of $\phi$
\be
 c_{\log}= - 0.04706\,,\hspace{.8cm}c_0=2.36-2 \log(\lambda/\sqrt{2\pi e})\,,\hspace{.8cm} c_1= - 2.39-2 \log(\lambda/\sqrt{2\pi e})\,. 
\ee

\noindent c) Center of $\pi$
\be
c_{\log}= -0.04703\,,\hspace{.8cm}c_0=-0.058-2 \log(\lambda/\sqrt{2\pi e})\,,\hspace{.8cm}c_1=- 0.014-2 \log(\lambda/\sqrt{2\pi e})\,,  
\ee
where $\lambda$ is the normalization constant in the formula for the Shannon entropy of the center (\ref{shannon1}).

Clearly, the area and constant terms suffer large changes with the change of algebra prescription. Of course, these non universal terms suffer other ambiguities in the continuum limit,  for example, they are not rotational invariant. It is remarkable the area term (and hence the full entropy) can easily turn to be negative due to the choice of the center prescription.  Negative entropies were found for non minimally coupled scalars and gauge fields in early calculations using the replica trick \cite{non-minimally}. The classical center hints to a natural explanation for these puzzling results.  

\begin{figure}[t]
\centering
\leavevmode
\epsfysize=5cm
\epsfbox{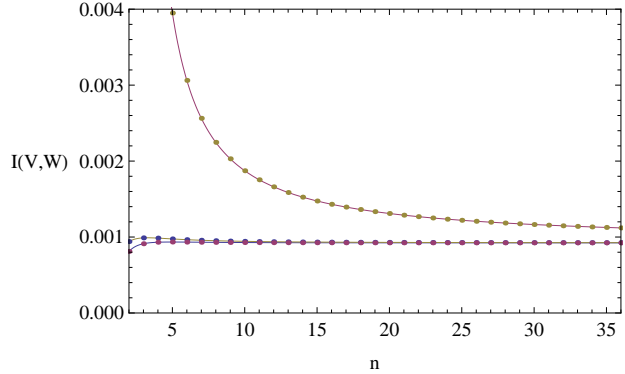}
\bigskip
\caption{Mutual information between two equal sized square regions separated by a distance equal to the squares sides. From top to bottom: Electric, trivial, and magnetic center choices.}
\label{figu8}
\end{figure}

\section{Entropy and mutual information for the gauge model}
We consider now the gauge model and calculate the entropies and mutual information for different geometries. Calculations are equivalently done in the dual model of a truncated scalar described in Section 2 using the tools developed in Section 4. 

First, we calculate the mutual information between two squares of equal sizes separated by a distance equal to the squares size for the three different algebra choices of figure (\ref{figu3}). The result is shown in figure (\ref{figu8}). As expected, the electric center has larger mutual information than the trivial center, and this is in turn larger than the magnetic center, in agreement with the monotonicity property of mutual information and the fact that the algebras in figure (\ref{figu3}) are ordered by inclusion.

We fit the mutual information as $I(V,W)=I_0+I_1 n^{-1}+I_2 n^{-2}+I_3 n^{-3}+ I_4 n^{-4}$. As claimed in \cite{gauge}, we find the same continuum limit for all of them, as shown in figure (\ref{figu8}) with
\bea
I^{E}_0&=&0.000923\,,\\
I_0^{T}&=&0.000924\,,\\
I_0^{M}&=&0.000924\,,
\eea
for the electric, trivial, and magnetic centers respectively. 

The contribution of the classical Shannon term to these universal numbers is shown in figure (\ref{centros}). We have that $H(V,W)$ goes to zero both for the electric and magnetic centers. It falls much faster for the magnetic center because it contains only one degree of freedom in contrast to the electric center which contains an area increasing number of degree of freedom. This also confirms general expectations that $H(V,W)$ has zero continuum limit \cite{gauge} because it is bounded above by the mutual information of regions on the boundary with lattice spacing (cutoff) width and fixed distance in the continuum interpretation. Hence, the Shannon term does not seem to have physical meaning in the continuum limit. This, together with the fact that all the superselection sectors for a given center give place to the same entropy (as shown in section 4) leads us to the conclusion that mutual information can be computed in the continuum limit from the reduced density matrix for just only one arbitrary sector, for example a fixed arbitrary normal electric field at the boundary in the electric center choice.      

\begin{figure}[t]
\centering
\leavevmode
\epsfysize=5cm
\epsfbox{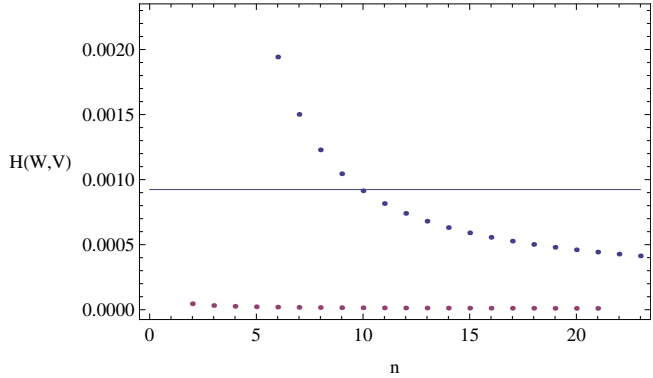}
\bigskip
\caption{Classical mutual information between the centers of two squares in the electric center (upper curve)
 and magnetic center (curve at the bottom). The horizontal line is the continuum limit of the full mutual information of the squares.}
\label{centros}
\end{figure}

Note the numerical value for the gauge field mutual information is around $35$ times smaller than the one corresponding one to the same two squares for the scalar field, $I=0.033$. In fact, the gauge model is (locally) identical to a subalgebra of the scalar field, the one generated by the gradient field. Hence, we expect the mutual informations for any two regions always satisfy 
\be
I_{\textrm{gauge}}(V,W)\le I_{\textrm{scalar}}(V,W)
\ee
 in $d=3$. 

We can learn more about the similitudes and differences between the scalar and gauge models by studying two limits on the mutual information.

\begin{figure}[t]
\centering
\leavevmode
\epsfysize=6cm
\epsfbox{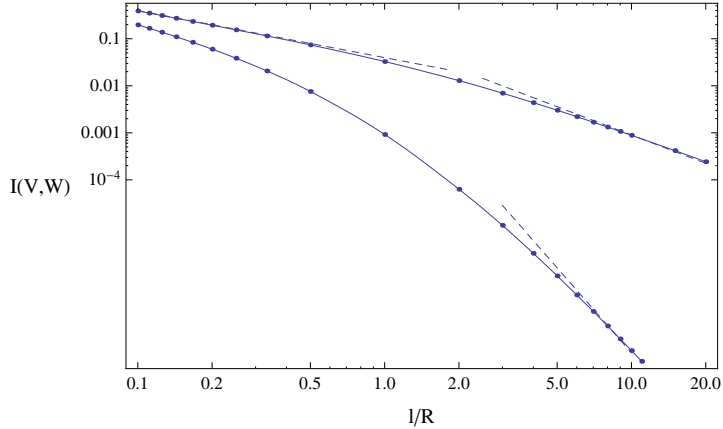}
\label{comparacion}
\bigskip
\caption{Log log plot of the mutual information for two squares of side $R$ separated by a distance $l$, as a function of $l/R$. The curve at the top is the mutual information for the scalar and the lower one in the mutual information of the gauge model. The dashed lines are asymptotic behaviors. For small $l/R$ we expect $I(V,W)\sim .0397 R/l$ for both models, while for large distances we expect $I(V,W)\sim (l/R)^2$ for the scalar and $I(V,W)\sim (l/R)^6$ for the Maxwell field.}
\end{figure}

First, when two regions with parallel phases approach each other, mutual information will diverge with the inverse of the distance. For two squares we have \cite{scal,review} 
\be
I(V,W)\sim k \frac{R}{l}+ ...  \hspace{2cm} \frac{l}{R}\ll 1\,.\label{cop}
\ee
This is an area term for the mutual information in the coincident limit. Now, we can argue that the constant coefficient $k$ must be the same for the scalar and the gauge fields, as follows.  The calculation of (\ref{cop}) for a scalar starts by realizing that in the small $l$ limit the term (\ref{cop}) is extensive in the direction of the coordinate $y$ parallel to the two nearby sides of the squares. Therefore, we can replace the two squares by two half-spaces for the sake of this computation. Then, we can compactify the space in a circle with large radius in the $y$ direction without changing the extensive part of the entropies (see \cite{review} for details). Decomposing the scalar into Fourier modes 
\be
\phi(x,y,t)= e^{i p_y y} \tilde{\phi}(x,t)\label{neq}
\ee
in the large direction, the mutual information $I(V,W)$ turns into a sum over the mutual informations of massive $1+1$ dimensional scalar fields, where the mass is produced by the momentum in the transverse direction, $m^2=p_y^2$. We have
\be
k_{\textrm{scalar}}=\frac{1}{\pi} \int_0^\infty dx \, C(x)\approx 0.0397 \,,\label{int}
\ee  
where $C(r m)=r d S(r,m)/dr$ is the entropic $C$-function of a massive scalar in $d=2$, and $S(r,m)$ is the entanglement entropy for an interval of size $r$ and a field of mass $m$.  Now, for the truncated scalar we can do the same calculation. For any mode with $p_y\neq 0$ in (\ref{neq}) the $1+1$ model produced by the truncated scalar is the same as the one of the full scalar. This is because for non-zero momentum the operators $\tilde{\phi}(x,t)$ in (\ref{neq}) belong to the truncated algebra as well, since the integral of these modes on the $y$ direction is exactly zero. Hence, the coefficient $k$ for the truncated scalar and the gauge field are given by the same integral (\ref{int}), differing only in a measure zero set at $m=0$, and we have
\be
k_{\textrm{gauge}}=k_{\textrm{scalar}}\approx 0.0397\,.
\ee

The second limit we want to look at is when the two regions $V$ and $W$ have large separations. For two squares this is the limit of $l/R\gg 1$, with $l$ the separation distance. In this case we have that mutual information falls as the square of the correlation function of the lowest dimension operator \cite{cardy3} (see also \cite{review} for the scalar case). This is $\phi$ for the scalar model and $F_{\mu\nu}$ for the gauge field. Thus, we expect
\bea
I_{\textrm{scalar}}\sim a_s \left(\frac{l}{R}\right)^{-2}\,, \hspace{2cm} \frac{l}{R}\gg 1\,,\label{ret1}\\
I_{\textrm{gauge}}\sim a_g \left(\frac{l}{R}\right)^{-6}\,,\hspace{2cm} \frac{l}{R}\gg 1\,.\label{ret2}
\eea
We can confirm these expectations for two squares in figure (\ref{comparacion}), where we have plotted mutual information against $l/R$. The short and long distance behavior nicely approach (\ref{cop}) and (\ref{ret1}), (\ref{ret2}). For two squares we can get the coefficients in (\ref{ret1}), (\ref{ret2}) approximately as $a_s\sim 0.09$, $a_g\sim 0.021$.

Summarizing, the mutual information for the scalar is always larger than the one for the gauge model. At short distances they have the same leading ultraviolet divergent term (area law) because they have the same ultraviolet modes, while at larger distances the scalar field has much larger mutual information. This is because the gauge field does not contain the scalar ``center of mass'' mode, $\sum_{i\in V} \phi_i$, which controls the largest share of mutual information for large distances. 

\subsection{Logarithmic term in the entropy}

Let us compute the logarithmic terms on the entropy for different choices of center. These are generally universal terms, and we have seen they do not depend on the center for the scalar field. Again we calculate the entropy for squares as in figure (\ref{figu3}) and fit with a function of the form $S=c_1 n+ c_{\log } \log(n)+c_{-1} n^{-1}+ c_{-2} n^{-2}+...$ with squares of size up to $35$ points. We get for the trivial center case (see figure \ref{figu7})
\bea
c_{\log}&=&0.4521\,,\\
c_{\log}&-&c^s_{\log}=0.4521+ 0.0472=.4993 \approx \frac{1}{2}\,,
\eea
where $c_{\log}^s$ is the logarithmic coefficient for a scalar. Surprisingly, for the gauge model we get $1/2$ plus the logarithmic coefficient for the scalar on the square. For the electric and magnetic centers we get similarly
\bea
c_{\log}^E&=& 0.4519\,,\\
c_{\log}^M&=& 0.4517\,.
\eea
Thus, the logarithmic term is independent of the center choice. We have also checked it is rotational invariant in the lattice (that is, it does not change for rotated squares).

\begin{figure}[t]
\centering
\leavevmode
\epsfysize=5.5cm
\epsfbox{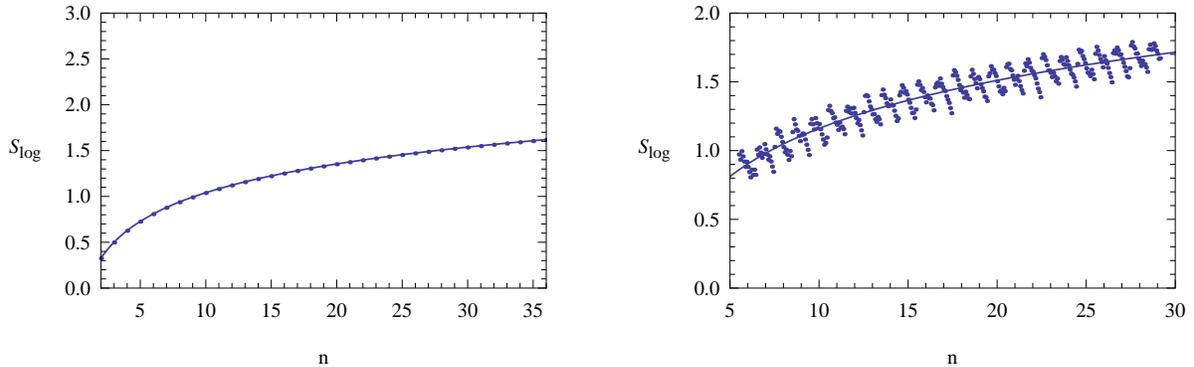}
\caption{Left panel: Entropy of a square of size $n$ where the linear term in a fit $S=c_1 n+c_0+s_{\log} \log(n)$ has been subtracted. The logarithmic term is shown with a solid line. Right panel: Entropy of circles of radius $n$ in the square lattice, computed with $1/10$ steps for the radius.  The linear term has been subtracted. Both figures are for a trivial center.}
\label{figu7}
\end{figure}

However, as we have pointed out, the entropies for algebras with continuum center are not well defined, in the sense that its classical contribution (\ref{class}) can vary with field normalizations. To understand the validity of these results for the logarithmic term we notice that a change in normalization by a factor $\lambda$ changes the entropy of the square by $-D_C\log(\lambda)$, where $D_C$ is the number of degrees of freedom in the center. This grows with the area in the electric center and is just one for the magnetic center. Hence, no changes in the logarithmic term are expected if we change the normalization by a factor independent of the number of points in the square. Changing the normalization by a factor depending on $n$ does not seem to be fair, in the sense that it would be a prescription for doing computations which includes information a priori on the object on which one wants to compute the entropy.  

In this regard, it is interesting to note that the classical entropy of the center for the electric choice does indeed give a non zero contribution to the logarithmic term (this is in fact a large fraction of the logarithmic term). Hence, even if the mutual information does not depend on the classical terms in the continuum limit, the logarithmic term in the entropy is sensible to the classical contributions. 

To discern how is this contribution related to the presence of angles in the square, we have computed the logarithmic term for several other shapes illustrated on figure \ref{config}. Here, we list the results for $c_{\log}$ for the different regions and compare with the $c^s_{\log}$ coefficient for the same regions in the full massless scalar theory (which is always a sum over the contributions of the different angles):

\noindent 
a)  Entropy of two equal squares separated by a distance of the size of the square:
\bea
c_{\log}&=& 0.9044\,,\\
c_{\log}&-&c^s_{\log}=c_{\log}+8\, s(\pi/2)= 0.999\approx 1\,.
\eea      

\noindent 
b) Square of size $3n$, with a centered square hole of size $n$:
\bea
c_{\log}&=& 0.4052\,,\\
c_{\log}&-&c^s_{\log}=c_{\log}+8\, s(\pi/2)= .4996 \approx \frac{1}{2}\,.
\eea

\begin{figure}[t]
\centering
\leavevmode
\epsfysize=2.5cm
\epsfbox{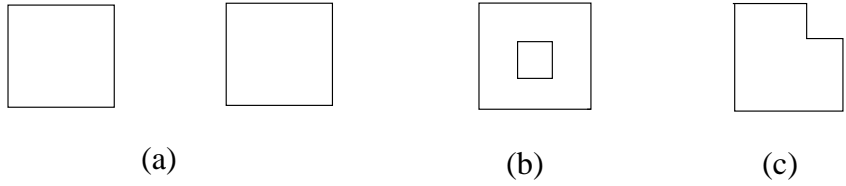}
\bigskip
\caption{Different regions with different topologies and different numbers of corners. Both of these features contribute to 
the logarithmic term for the gauge model.}
\label{config}
\end{figure}

\noindent
c) Square of size $3n$ with a removed corner square of size $n$:
\bea
c_{\log}&=& 0.4294\,,\\
c_{\log}&-&c^s_{\log}=c_{\log}+6\, s(\pi/2)=0.498\approx \frac{1}{2}\,.
\eea

In figure (\ref{figu7}) we also show the logarithmic term in the entropies of circles in the square lattice. In this case, no angle term is expected and we get $S_{\log}=0.504 \log(n)$ for circles of radius up to $n=30$. Of course, we have considerable noise for circles, but still the logarithmic term is clearly seen and can be extracted from the data with good precision. 

 From the above results and many other checks, including triangular regions with angles different from $\pi/2$, we conclude the general form of the logarithmic coefficient is 
\begin{equation}
c_{\log}= \frac{N_c}{2}+c_{\log}^s=\frac{N_c}{2}-\sum_{v} s(\theta_v)\,, 
\end{equation}
where $N_c$ is the number of connected components in the region, and $s(\theta_v)$ is exactly the same function giving the logarithmic coefficient due to the angles in the full scalar theory.  The presence of the same ultraviolet logarithmic term depending on the angles as in the scalar case can be explained by the same argument as for the coefficient $k$ in the mutual information (previous subsection). This is because the coefficient $s(\theta)$ is computed by dimensional reduction in spherical symmetry \cite{ang} and all relevant dimensionally reduced modes for the scalar coincide with the ones for the truncated scalar.

 On the other hand, the new contribution is topological and "counts" the number of components of the region independently of the shape. This contribution to the entropy is rather puzzling in two respects. The first one is that this is a term proportional to $\log(R/\epsilon)$ and hence apparently has an ultraviolet origin, but at the same time it does not look local on the boundary on geometrical grounds.\footnote{A dimensionless quantity which is a local integral on the boundary is
 $
 -\frac{1}{2 \pi}\int_{\partial V} ds \,\gamma(s)=2 N_c-N_b
 $, 
 where $\gamma(s)=\eta(s).\frac{d^2(x(s))}{ds^2}$ is the local oriented curvature of the boundary curve, $x(s)$ parametrizes the curve by length, and $\eta(s)$ is the outwards pointing normal unit vector at the boundary. Here $N_b$ is the number of disjoint boundaries of the region.  It is interesting to note this means the number of connected components we find in the logarithmic coefficient is then equivalent, up to local terms, to the number of boundaries \cite{tarum}. This resembles the topological entanglement entropy dependence on the region. We thank Tatsuma Nishioka for this comment.}  However, it is important to realize that this term is only apparently ultraviolet, and no short distance entanglement consequence of this term can be seen in the universal mutual information. In general we expect that divergent terms in the entropy can be re-obtained, in a regularization independent way, using the mutual information between the region $V$ and an external region $W$ that surrounds $V$, in the limit of small distance $\delta$ between these two regions. For example, for the scalar we have
 \be
 I(V,W)= k\frac{R}{\delta} + 2 \log(\delta) \sum_v s(\theta_v)+ {\cal O}(\delta^0)\,,
 \ee 
 where $R$ is the region perimeter.  
For the truncated scalar we should have the same formula, with the same coefficients, plus the term $-\log(\delta) N_c$. However, this last term is clearly impossible, since the mutual information of the truncated scalar is bounded above by the one of the scalar, and this term would violate this inequality for small enough $\delta$. A direct check of this on the lattice is difficult because we have to go to the small $\delta$ limit but having first reached the continuum limit. We will say a bit more about this topic and its possible relation with the $c$-theorem in $d=3$ \cite{headrick,kleba} in the discussion section. 
 
 The second puzzle is the positive sign of the logarithmic term $S_{\log}=N_c/2 \log(R/\delta)$. This is rather startling because the negative sign for the logarithmic terms coming from the vertices in $d=3$ (see for example formula (\ref{yapeyu})) is a general property imposed by strong subadditivity.  We left for the next section the solution of the paradox on how strong subadditivity is preserved against all odds for this entropy function.

\section{Strong subadditivity and algebras}
\label{ssassa}
Historically, SSA was first proposed in order to prove the stability of the matter through concavity of the entropy function in translation invariant systems \cite{conjecture-ssa}. It was formulated for the case of Hilbert spaces that can be written as tensor products. Consider $\mathcal{H}={\mathcal{H}}^1 \otimes \mathcal{H}^2 \otimes\mathcal{H}^{3}$, then \cite{Lieb} 
\be
S(\rho^{123})+S(\rho^2)\leq S(\rho^{12})+S(\rho^{23})\,,\label{termi}
\ee
where $S(\rho^{12})=-\textrm{Tr}_{\mathcal{H}^{12}} \rho^{12}\log\rho^{12}$, etc..
In general this is supposed to give place to the geometric inequality 
\be
S(A)+S(B)-S(A \cup B)-S(A \cap B)\label{ssageo}
\ee
for the entropy of regions. 
However, we have to be more careful in the gauge field case, since the tensor product decomposition may not be possible. 

\begin{figure}[t]
\centering
\leavevmode
\epsfysize=4.5cm
\epsfbox{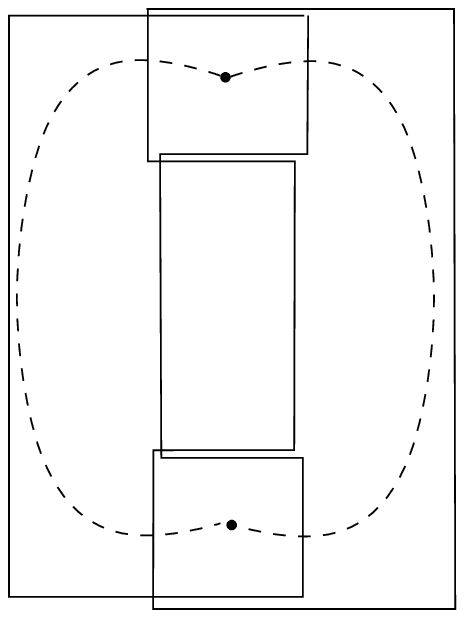}
\bigskip
\caption{Two horseshoe like regions intersect each other. The operators $\sum \hat{\phi}$ summed along each of the two lines joining the two marked points are equal due to the constraint relation (telling the sum of $\hat{\phi}$ along any closed curve is zero).}
\label{figu91}
\end{figure}

The Maxwell field in $2+1$ dimensions gives a clear example. Suppose we have two horseshoe-like regions $A$ and $B$, intersecting in two disjoint squares, and having an annulus shaped union as in figure (\ref{figu91}) and (\ref{figu9}). According to the previous discussion, the entropy for these regions has some divergent terms: An area term, a logarithmic term which is a sum over the different angles, and a  logarithmic term proportional to the number of connected components. The first two terms cancel in the combination (\ref{ssageo}), while the last one would give 
\be
S(A)+S(B)-S(A \cup B)-S(A \cap B)=- \frac{1}{2}\log(R/\epsilon)+ \textrm{const}< 0\,,\label{hi}
\ee
for some macroscopic size $R$, violating strong subadditivity. 

The first problem with this argument is again the ambiguities in the assignation of algebras to regions. What are the algebras corresponding to the different regions in (\ref{ssageo}) and how they are related to each other? With the usual assignations of the algebras for regions there is no way to choose the local algebras for the regions in (\ref{hi}) to eliminate the remaining logarithmic term. This term comes with the wrong sign, and SSA in this form does not hold. 

As it happens, what is wrong here is the idea that strong subadditivity can be applied to algebras assigned to regions as in the left hand side of (\ref{hi}). Instead of that, the set operations of intersection and union  have to be applied to the algebras themselves. We propose the inequality
\be
F({\cal  A}_A,{\cal  A}_B)=S({\cal  A}_A)+S({\cal  A}_B)- S({\cal  A}_{A}\vee {\cal  A}_{B} )-S({\cal  A}_{A}\wedge {\cal  A}_{B} )\ge 0\,.\label{hjkl}
\ee    
Here ${\cal  A}_{A}\wedge {\cal  A}_{B}$ is the intersection of algebras (which is another algebra) and ${\cal  A}_{A}\wedge {\cal  A}_{B}=({\cal  A}_{A}\cup {\cal  A}_{B})^{\prime\prime}$ is the algebra generated by the two (the smallest algebra containing the two). In the present form, strong subadditivity holds for the two horseshoe regions. The reason is that the intersection of the two algebras of the horseshoe regions is the algebra of a region with two square components  but there is an additional extra long distance link $\phi_1-\phi_2$ on the algebra of the intersection (in the truncated scalar formulation, see figure \ref{figu9}), where $\phi_1$ and $\phi_2$ are any two field operators localized in each of the two separated squares forming the intersection. This is because  the two strings of fields shown in figure (\ref{figu91}) are the same operator due to the constraint equations. Therefore, this string belongs both to ${\cal  A}_A$ and ${\cal  A}_B$. 
In the gauge formulation the long link represents the global flux of electric field normal to the line. 
Because of this extra link, the intersection effectively looses one component, and in fact the calculation of the logarithmic term gives the same result than for a one component region. As a consequence, divergent pieces cancel exactly in (\ref{hjkl}), and it turns out the combination is positive.\footnote{Note that for theories with charges the intersection does not contain the flux of the electric field on a line joining the two regions. This is because the constraint has an additional piece of the charge contained inside the closed loop which effectively decouples the two fluxes on the two horseshoe algebras. According to the present discussion, this means no divergent term proportional to $\log(R/\epsilon)$ can appear for charged theories.} This is shown in figure \ref{figu10} where this quantity is calculated for $A$ and $B$ with the magnetic, trivial and electric center algebras. As shown in the figure \ref{figu10}, all cases have the same continuum limit. Fitting the curves we obtain the constant limit
\footnote{Note in the electric center case the center is formed by boundary links, but the center of the union has an additional large distance link connecting the two boundaries.}
\bea
F({\cal  A}_A,{\cal  A}_B)&=&0.08583, \,\,\,\,\textrm{electric center}\,,\\
F({\cal  A}_A,{\cal  A}_B)&=&0.08575, \,\,\,\,\textrm{trivial center}\,,\\
F({\cal  A}_A,{\cal  A}_B)&=&0.08579,\,\,\,\, \textrm{magnetic center}\,.
\eea

For SSA in the usual form (\ref{termi}) the relation (\ref{hjkl}) is clear since all algebras involved are in tensor products. The SSA relation for algebras with center but still in tensor products is shown in \cite{petz}. But the crucial point in this example, is that given the algebras of ${\cal  A}_A$ and ${\cal A}_B$ in the gauge model, with any of the possible choices, we get $F({\cal  A}_A,{\cal  A}_B)\ge 0$, even if the tensor product decompositions in three algebras as in (\ref{termi}) cannot be achieved.

\begin{figure}[t]
\centering
\leavevmode
\epsfysize=6cm
\epsfbox{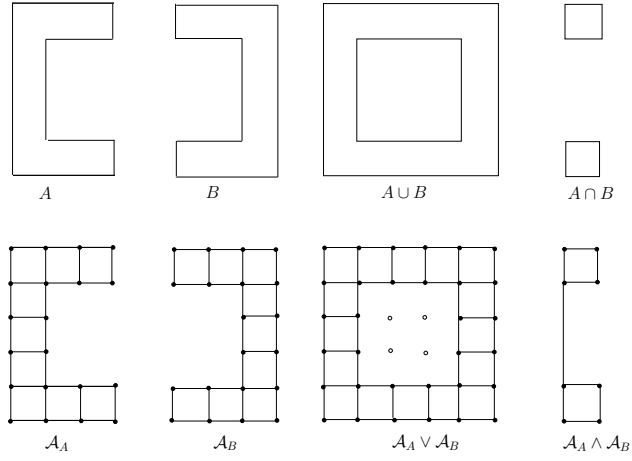}
\bigskip
\caption{The figures at the top show two horseshoe regions $A$ and  $B$, and their union and intersections $A\cup B$, $A\cap B$ (once $A$ and $B$ became superimposed as in figure \ref{figu91}). The figures at the bottom show the algebras ${\cal  A}_A$, ${\cal  A}_B$, ${\cal  A}_{A}\vee {\cal  A}_{B} $ and ${\cal  A}_{A}\wedge {\cal  A}_{B}$ in the truncated scalar model (magnetic center in this example). The intersection of the algebras contains an additional non-local link with respect to the algebra corresponding to the intersection $A\cap B$. This non-local link crossing between the two components makes the intersection of the algebras an effective one component set.}
\label{figu9}
\end{figure}

We could not find in the literature the statement (\ref{hjkl}) for strong subadditivity and naturally wonder about the conditions for its validity.  It is clear it is not valid in full generality as shown by the example of two canonical variables, where ${\cal  A}_A=\{q_1,p_1\}$, ${\cal  A}_B=\{q_1+\epsilon q_2,p_2+\epsilon p_2\}$, ${\cal  A}_{A}\vee {\cal  A}_{B}=\{q_1,p_1,q_2,p_2\} $ and ${\cal  A}_{A}\wedge {\cal  A}_{B}=1$. In this case if the first pair of canonical variables are in a pure state decoupled from the second pair, and this second pair has large entropy, we have $S({\cal  A}_A)\sim S({\cal  A}_B)\sim 0$, $S({\cal  A}_{A}\vee {\cal  A}_{B})$ is large and $S({\cal A}_{A}\wedge {\cal  A}_{B})=0$. Then (\ref{hjkl}) does not hold. Another example is given by time like separated regions in QFT. We want to exclude these cases.  

A possible condition is as follows. The case of tensor products (\ref{termi}) is characterized algebraically by the fact that the following ``commutator'' algebra\footnote{This is the terminology in the theory of orthomodular lattices \cite{orthomodular}, which has strong connections with von Neumann algebras and the relations between causal regions in spacetime \cite{causal2,causal3}. In the orthomodular lattice of causal regions the relation analogous to (\ref{milder}) means that the causal sets $A$ and $B$ have a common Cauchy surface \cite{causal3}.} is trivial
\be
{\cal C}({\cal  A}_A,{\cal  A}_B)\equiv({\cal  A}_A\vee {\cal  A}_B)\wedge ({\cal  A}_{A}\vee {\cal  A}_{B}^\prime)\wedge 
({\cal  A}_A^\prime \vee {\cal  A}_B)\wedge ({\cal  A}_{A}^\prime \vee {\cal  A}_{B}^\prime)=1 \label{milder}
 \,.
\ee
This is not the case of our lattice algebras nor of the above example with two canonical pairs. However, for the example with two canonical variables we get
\be
{\cal C}({\cal  A}_A,{\cal  A}_B)={\cal  A}_A\vee {\cal  A}_B\,.
\ee
For the algebras of our lattice simulations, we get a milder modification of (\ref{milder}), namely that 
\be
{\cal C}({\cal  A}_A,{\cal  A}_B)=Z \label{milder1}
 \,,
\ee
with $Z$ a commuting algebra. For the lattice algebras, $Z$ is in fact generated by operators localized at the boundaries of $A$ and $B$ which commute to each other (however they do not necessarily belong to the center of ${\cal A}_A$ and ${\cal A}_B$, which can be full algebras). Hence, in certain sense, the modification of the conditions for SSA is again a detail at the boundaries, and in  algebraic terms the condition (\ref{milder1}) should be read as equivalent to the condition for regions $A$ and $B$ (as represented in the model by their algebras) to be lying in the same Cauchy surface, or not to be time-like to each other.     
 Hence, we are tempted to propose (\ref{milder1}) as the mathematical condition of validity of SSA in the form (\ref{hjkl}). 

\begin{figure}[t]
\centering
\leavevmode
\epsfysize=5cm
\epsfbox{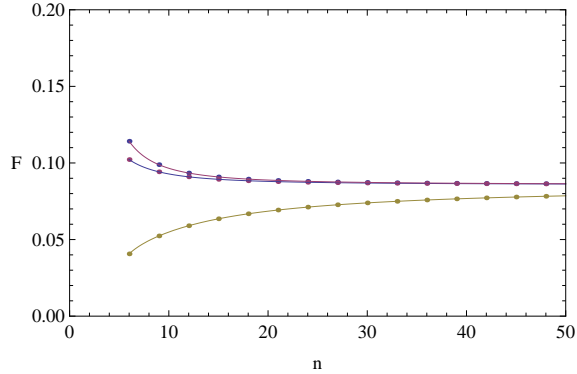}
\bigskip
\caption{$F({\cal  A}_A,{\cal  A}_B)=S({\cal  A}_A)+S({\cal  A}_B)- S({\cal  A}_{A}\vee {\cal  A}_{B} )-S({\cal  A}_{A}\wedge {\cal  A}_{B})$ for the configuration of figure \ref{figu9} as a function of the overall size in the lattice (here the side $n$ of the big square for the union of the two regions). The curves are for algebras with magnetic, trivial, and electric center (from top to bottom).}
\label{figu10}
\end{figure}

\section{Final comments}
We have shown that it is possible to make numerical computations of entanglement entropy for free gauge fields in a lattice with the same computational complexity as for a scalar field.

Our main general conclusion is that ambiguities on the algebra choice in the gauge model disappear in the continuum limit for the same terms in the entropy that are universal for other fields. We have also seen the classical Shannon term is not relevant for the continuum limit of mutual information. In fact, at least for the free models, the calculation of quantities of the continuum, can as well be reduced to the quantum term on a single arbitrary sector in the central decomposition of local algebras with center.  
 
We were compelled to propose a form of strong subadditivity which is written in algebraic terms. We conjecture some particular condition for this relation to be valid, with generalizes the usual strong subadditive inequality for tensor products. Note that the proof of the existence of an entropy density for translational invariant states \cite{conjecture-ssa}, which was the historical motivation to introduce SSA, would need of this enlarged form of SSA to be freely applied to gauge fields.

\subsubsection*{Topological entanglement entropy}
The discussion of strong subadditivity for the two horseshoe like regions in section (\ref{ssassa}) has close connections with the topological entanglement entropy \cite{wen,kitaev}. This geometric arrangement was used to define the topological entanglement entropy by Levin and Wen,
\be
S(A)+S(B)-S(A\cup B)-S(A\cap B)=\gamma_{\textrm{topo}}
\ee  
for gapped models in $2+1$ dimensions. However, as we have seen, the quantity on the left hand side, has to be defined in an algebraic way as $F({\cal A}_A, {\cal A}_B)$. We compute this quantity in a topological $Z_2$ gauge model following our previous calculations \cite{gauge} (see also \cite{zana,Polikarpov}). We find that for all prescriptions of the local algebras we get $F({\cal A}_A, {\cal A}_B)=0$, essentially because the number of degrees of freedom in the centers of $\cal{A}_A$ and $\cal{A}_B$ are equal to the ones in $\cal{A}_A\vee\cal{A}_B$ and $\cal{A}_A\wedge\cal{A}_B$. This suggests topological entanglement entropy needs UV degree of freedom to be well defined (see however \cite{pibe}). That is, we need some degree of freedom in the UV which allow the continuum limit to be taken. If the topological model is a long distance limit of some other theory at the UV we expect $F({\cal A}_A, {\cal A}_B)$ to give the topological entropy for large regions.  

\begin{figure}[t]
\centering
\leavevmode
\epsfysize=5cm
\epsfbox{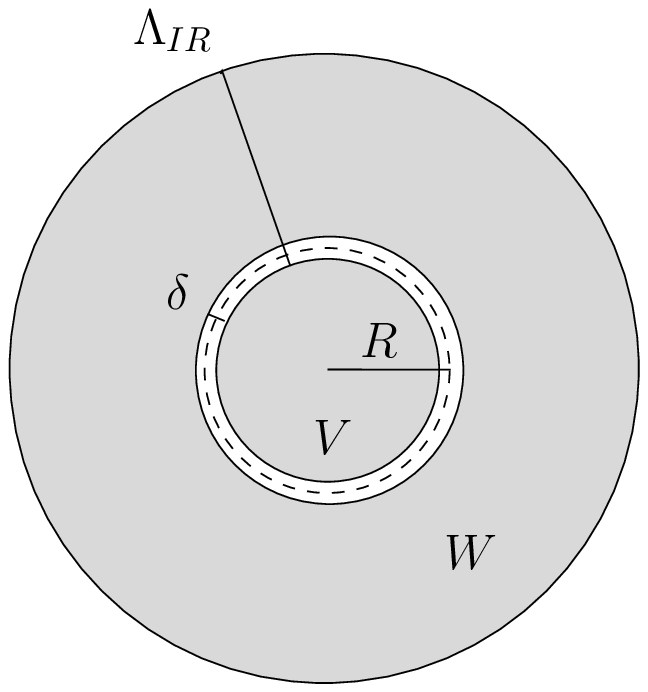}
\bigskip
\caption{The region $W$ encircles $V$ at a distance $\delta$. The large radius $\Lambda_{IR}$ is used as an infrared cutoff.}
\label{ultima}
\end{figure}

\subsubsection*{Logarithmic term and c-theorem}

We have found a logarithmic term $1/2 \log(R/\epsilon)$ in the entropy with topological coefficient. However, this term is not reproduced by the mutual information, see figure \ref{ultima}. In this calculation, we have the region $V$ surrounded by the region $W$ at a distance $\delta$, acting as an UV cutoff, and $W$ has a maximal size $\Lambda_{IR}$ acting as an infrared cutoff. As $\delta\rightarrow 0$, we found that there cannot be any $\sim -\log(\delta)$ term in the mutual information, since mutual information of the gauge model is bounded above by the one of the scalar, which does not have such term (while it has the same coefficient for the area term). We expect however, the behavior $1/2 \log(R)$ to be reproduced in the mutual information at least as the infrared size $\Lambda_{IR}$ goes to infinity. This is the geometric regularization of (twice the) entropy by mutual information \cite{mutureg}. Since the only finite available sizes in this construction are $R$ and $\Lambda_{IR}$ we expect to have a term
\be
I(V,W)_{\log}\sim \log(R/\Lambda_{IR})\,.\label{nicely}
\ee
Note that as $\Lambda_{IR} \gg R$ this term is negative. Hence, we have to take $\delta \ll R,\Lambda_{IR}$ for this to make sense. Taking the limit of large $\Lambda_{IR}$ at fixed $\delta$, this term cannot appear because it would make mutual information negative.  

Eq. (\ref{nicely}) would nicely fit with the results of \cite{headrick}. There, the authors computed the entropy for a disc and a compactified Maxwell field and found a logarithmic term $1/2 \log(R g^2)$, with $g$ the gauge coupling constant, in the entanglement entropy, in the limit of $R g^2\ll 1$. This can be thought as a large positive term in the $c$-function $R S^\prime(R)-S(R)$ of the $d=3$ c-theorem \cite{nunu} (see also \cite{liu}).  As pointed out in \cite{kleba} a logarithmically divergent term in the $c$-function in the limit of uncompactified  Maxwell field ($R g^2\ll 1$) is necessary for the validity of the c-theorem due to a possible running of the Maxwell theory at the UV to a Chern Simmons theory in the infrared.  

This is also consistent with the observation \cite{mutual} that the correct definition of the c-function is in terms of the constant term in the expansion of mutual information in powers of $\delta$, $I(R-\delta/2,R+\delta/2) \sim 2 c_1 \frac{R}{\delta}-2 c_0+...$.
The large $\Lambda_{IR}$ limit has to be taken after the small $\delta$ limit (though in general we expect these limits to commute).  Note $c_0$ cannot be directly extracted from the entropy by the formula $c_0=R S^\prime(R)-S(R)$ in this particular model, because it contains a large but negative term.  

Unfortunately, to check these expectations with some precision in the square lattice seems to be difficult. We left for a future work the relevant calculation in a radial lattice.

\section*{Acknowledgements}
We thank the Institute for Advanced Study for hospitality and financial support during the realization of this project. 
It is a pleasure to thank Matthew Headrick, Igor Klebanov, Tatsuma Nishioka, Benjamin Safdi, for discussions. 
This work was partially supported by CONICET, CNEA and Universidad Nacional de Cuyo, Argentina.

\end{document}